\documentstyle[12pt]{article}
\begin{document}

\title{\bf{Reconstructing interacting entropy-corrected holographic scalar field models of dark energy in non-flat universe}}

\author{K. Karami$^{1,2}$\thanks{E-mail: KKarami@uok.ac.ir}
,~M.S. Khaledian$^{1}\thanks{E-mail: MS.Khaledian@uok.ac.ir}$
,~Mubasher Jamil$^{3}$\thanks{E-mail:
mjamil@camp.nust.edu.pk}\\\\
$^{1}$\small{Department of Physics, University of Kurdistan,
Pasdaran St., Sanandaj, Iran}\\$^{2}$\small{Research Institute for
Astronomy $\&$ Astrophysics of Maragha (RIAAM), Maragha,
Iran}\\$^{3 }$\small{Center for Advanced Mathematics and Physics
(CAMP), National University} \\\small{of Sciences and Technology
(NUST), Islamabad, Pakistan}}

\maketitle

\begin{abstract}
Here we consider the entropy-corrected version of the holographic
dark energy model in the non-flat universe. We obtain the equation
of state parameter in the presence of interaction between dark
energy and dark matter. Moreover, we reconstruct the potential and
the dynamics of the quintessence, tachyon, K-essence and dilaton
scalar field models according to the evolutionary behavior of the
interacting entropy-corrected holographic dark energy model.
\end{abstract}
\noindent{PACS numbers: 95.36.+x, 04.60.Pp}\\
\clearpage
\section{Introduction}
The present acceleration of the universe expansion has been well
established through numerous and complementary cosmological
observations \cite{Riess}. A component which is responsible for this
accelerated expansion usually dubbed ``dark energy'' (DE). However,
the nature of DE is still unknown, and people have proposed some
candidates to describe it (for a good review see
\cite{Padmanabhan,Copeland} and references therein).

The holographic DE (HDE) is one of interesting DE candidates which
was proposed based on the holographic principle \cite{Horava}.
According to the holographic principle, the number of degrees of
freedom of a physical system should scale with its bounding area
rather than with its volume \cite{Hooft} and it should be
constrained by an infrared cut-off \cite{Cohen}. By applying the
holographic principle to cosmology, one can obtain the upper bound
of the entropy contained in the universe \cite{Fischler}.
Following this line, Li \cite{Li} suggested the following
constraint on its energy density
$\rho_{\Lambda}\leq3c^2M_P^2L^{-2}$, the equality sign holding
only when the holographic bound is saturated. In this expression
$c$ is a numerical constant, $L$ denotes the IR cut-off radius and
$M_P=(8\pi G)^{-1/2}$ is the reduced Planck Mass. The HDE models
have been studied widely in the literature
\cite{Enqvist,Elizalde2,Guberina1,Guberina2}. Obviously, in the
derivation of HDE, the black hole entropy $S_{\rm BH}$ plays an
important role. As is well known, usually, $S_{\rm BH} = A/(4G)$,
where $A\sim L^2$ is the area of horizon. However, in the
literature, this entropy-area relation can be modified to
\cite{modak}
\begin{equation}
S_{\rm
BH}=\frac{A}{4G}+\tilde{\alpha}\ln{\frac{A}{4G}}+\tilde{\beta},\label{MEAR}
\end{equation}
where $\tilde{\alpha}$ and $\tilde{\beta}$ are dimensionless
constants of order unity. These corrections can appear in the black
hole entropy in loop quantum gravity (LQG) \cite{HW}. They can also
be due to thermal equilibrium fluctuation, quantum fluctuation, or
mass and charge fluctuations (for review see \cite{HW} and
references therein). Using the corrected entropy-area relation
(\ref{MEAR}), the energy density of the entropy-corrected HDE
(ECHDE) can be obtained as \cite{HW}
\begin{equation}
\rho_{\Lambda}=3c^2M_P^2L^{-2}+\alpha L^{-4}\ln(M_P^2L^{2})+\beta
L^{-4},\label{ECHDE}
\end{equation}
where $\alpha$ and $\beta$ are dimensionless constants of order
unity. In the special case $\alpha=\beta=0$, the above equation
yields the well-known HDE density. Since the last two terms in Eq.
(\ref{ECHDE}) can be comparable to the first term only when $L$ is
very small, the corrections make sense only at the early stage of
the universe. When the universe becomes large, ECHDE reduces to the
ordinary HDE \cite{HW}.

Reconstructing the holographic and agegraphic scalar field models
of DE is one of interesting issue which has been investigated in
the literature \cite{Zhang11,Karami,Wu}. The holographic and the
agegraphic DE models are originated from some considerations of
the features of the quantum theory of gravity. On the other hand,
the scalar field models (such as quintessence, tachyon, K-essence
and dilaton) are often regarded as an effective description of an
underlying theory of DE \cite{Wu}. The scalar field models can
mimic cosmological constant at the present epoch and can give rise
to other observed values of the equation of state parameter
$\omega$ (recent data indicate that $\omega$ lies in a narrow
strip around $\omega=\omega_{\Lambda}= -1$ and is consistent with
being below this value) \cite{Ali}. They can also alleviate the
fine tuning and coincidence problems \cite{Ali}. Therefore it
becomes meaningful to reconstruct the scalar field models from
some DE models possessing some significant features of the LQG
theory, such as ECHDE and entropy-corrected agegraphic DE (ECADE)
models.

An interesting feature of entropy-corrected DE is that it permits
successive acceleration-deceleration phase transitions. Moreover
the cosmic coincidence problem is resolved and the universe
eventually tends to de Sitter expansion \cite{mohseni}. Here our
aim is to investigate the correspondence between the
entropy-corrected version of the interacting HDE model with the
quintessence, tachyon, K-essence and dilaton scalar field models
in the non-flat universe. These correspondences are essential to
understand the connection of various scalar field models of DE
with the ECHDE. This paper is organized as follows. In Section 2,
we obtain the equation of state parameter for the interacting
ECHDE model in a non-flat universe. In Sections 3-6, we suggest a
correspondence between the interacting ECHDE and the quintessence,
tachyon, K-essence and dilaton scalar field models in the presence
of a spatial curvature. We reconstruct the potentials and the
dynamics for these scalar field models, which describe accelerated
expansion of the universe. Section 7 is devoted to conclusions.
\section{Interacting ECHDE and DM in non-flat universe}
Within the framework of the standard FRW cosmology,
\begin{equation}
{\rm d}s^2=-{\rm d}t^2+a^2(t)\left(\frac{{\rm
d}r^2}{1-kr^2}+r^2{\rm d}\Omega^2\right),\label{metric}
\end{equation}
for the non-flat FRW universe containing the ECHDE and DM, the
first Friedmann equation takes the form
\begin{equation}
{\textsl{H}}^2+\frac{k}{a^2}=\frac{1}{3M_P^2}~
(\rho_{\Lambda}+\rho_{\rm m}),\label{eqfr}
\end{equation}
where $k=0,1,-1$ represent a flat, closed and open FRW universe,
respectively. Also $\rho_{\Lambda}$ and $\rho_{\rm m}$ are the
energy density of ECHDE and DM, respectively. Observational
evidences have implied that our universe is not a perfectly flat
universe and that it possesses a small positive curvature
\cite{Bennett}. Besides, as usually believed, an early inflation
era leads to a flat universe. This is not a necessary consequence
if the number of e-foldings is not very large \cite{Huang}.
Additionally the parameter $\Omega_k$ (discussed below) represents
the contribution in the total energy density from the spatial
curvature and is constrained as $-0.0175<\Omega_k<0.0085$ with
$95\%$ confidence level by current observations \cite{water}. It
has been shown that a non-zero positive curvature parameter $k$
allows for a bounce, thereby preventing the cosmic singularities
without violating the null energy condition $\rho+p\geq0$
\cite{paris}.

From Eq. (\ref{eqfr}), we can write
\begin{equation}
\Omega_{\rm m}+\Omega_{\Lambda}=1+\Omega_{k},\label{eq10}
\end{equation}
where we have used the following definitions
\begin{equation}
\Omega_{\rm m}=\frac{\rho_{\rm m}}{\rho_{\rm cr}}=\frac{\rho_{\rm
m}}{3M_P^2H^2},~~~~~~\Omega_{\rm
\Lambda}=\frac{\rho_{\Lambda}}{\rho_{\rm
cr}}=\frac{\rho_{\Lambda}}{3M_P^2H^2},~~~~~~\Omega_{k}=\frac{k}{a^2H^2}.
\label{eqomega}
\end{equation}
The recent observational evidence provided by the galaxy cluster
Abell A586 supports the interaction between DE and DM
\cite{Bertolami8}. This motivates us to consider the interaction
between ECHDE and DM. Hence $\rho_{\Lambda}$ and $\rho_{\rm m}$ do
not conserve separately and the energy conservation equations for
ECHDE and DM are
\begin{equation}
\dot{\rho}_{\Lambda}+3H(1+\omega_{\Lambda})\rho_{\Lambda}=-Q,\label{eqpol}
\end{equation}
\begin{equation}
\dot{\rho}_{\rm m}+3H\rho_{\rm m}=Q,\label{eqCDM}
\end{equation}
where following \cite{Kim06}, we choose $Q=\Gamma\rho_{\Lambda}$
as an interaction term and
$\Gamma=3b^2H(\frac{1+\Omega_{k}}{\Omega_{\Lambda}})$ is the decay
rate of the ECHDE component into DM with a coupling constant
$b^2$. Although this expression for the interaction term may look
purely phenomenological but different Lagrangians have been
proposed in support of it \cite{Tsujikawa}.

Note that in Eq. (\ref{ECHDE}), taking $L$ as the size of the
current universe, for instance, the Hubble scale, the resulting
energy density is comparable to the present day DE. However, as
found by Hsu \cite{Hsu}, in that case, the evolution of the DE is
the same as that of DM (dust matter), and therefore it cannot
drive the universe to accelerated expansion. The same appears if
one chooses the particle horizon of the universe as the length
scale $L$ \cite{Li}. To obtain an accelerating universe, Li
\cite{Li} proposed that for a flat universe, $L$ should be the
future event horizon $R_{\rm h}$ and Huang and Li \cite{Huang}
argued that for the non-flat case, the IR cut-off $L$ should be
defined as
\begin{equation}
L=a\frac{\sin n(\sqrt{|k|}y)}{\sqrt{|k|}},\label{L}
\end{equation}
where
\begin{equation}
y=\frac{R_{\rm h}}{a}=\int_t^\infty \frac{{\rm d} t}{a}=\int_0^{r}
\frac{{\rm d}r}{\sqrt{1-kr^2}}.\label{y}
\end{equation}
Here $R_{\rm h}$ is the radial size of the event horizon measured in
the $r$ direction and $L$ is the radius of the event horizon
measured on the sphere of the horizon \cite{Huang}. For a flat
universe, $L=R_{\rm h}$. The last integral in Eq. (\ref{y}) has the
explicit form as
\begin{equation}
\int_0^{r} \frac{{\rm d}r}{\sqrt{1-kr^2}}=
\frac{1}{\sqrt{|k|}}\sin n^{-1}\Big(\sqrt{|k|}r\Big)=\left\{
\begin{array}{ll}
\sin^{\rm -1}r,& k=1, \\
r,& k=0, \\
\sinh^{\rm -1}r,&k=-1. \\
\end{array}
\right.\label{horizon}
\end{equation}
From definition $\rho_{\Lambda}=3M_P^2H^2\Omega_{\Lambda}$ and
using Eq. (\ref{ECHDE}), we get
\begin{equation}
L=\frac{c}{H}\left(\frac{\gamma_c}{\Omega_{\Lambda}}\right)^{1/2},\label{HL}
\end{equation}
where
\begin{eqnarray}
\label{gamma-parameter1} \gamma_c = 1 +
\frac{1}{3c^2{M_P^2}L^2}\Big[\alpha\ln{({M_P^2}{L}^2)}
+\beta\Big].\label{gamma n}
\end{eqnarray}
Taking time derivative of Eq. (\ref{L}) and using (\ref{HL}) yields
\begin{equation}
\dot L=c\left(\frac{\gamma_c}{\Omega_{\Lambda}}\right)^{1/2}-\cos
n\Big(\sqrt{|k|}y\Big),\label{Ldot}
\end{equation}
where
\begin{equation}
\cos n\Big(\sqrt{|k|}y\Big)=\left\{
\begin{array}{ll}
\cos y,& k=1, \\
1,& k=0, \\
\cosh y,&k=-1. \\
\end{array}
\right.\label{cosn}
\end{equation}
Taking time derivative of Eq. (\ref{ECHDE}) and using (\ref{HL}) and
(\ref{Ldot}), one can obtain

\begin{eqnarray}
\dot{\rho_\Lambda}=\left(\frac{2H\rho_{\Lambda}}{\gamma_c}\right)
\left[1-2\gamma_c+\frac{\alpha
H^2}{3c^2M_P^2}\left(\frac{\Omega_{\Lambda}}{c^2\gamma_c}\right)\right]
\left[1-\left(\frac{\Omega_\Lambda}{c^2\gamma_c}\right)^{1/2}\cos
n\Big(\sqrt{|k|}y\Big)\right].\label{ro dot}
\end{eqnarray}

Substituting Eq. (\ref{ro dot}) in (\ref{eqpol}) gives the equation
of state (EoS) parameter of the interacting ECHDE as
\begin{eqnarray}
\omega_\Lambda=-1-b^2\left(\frac{1+\Omega_k}{\Omega_\Lambda}\right)
~~~~~~~~~~~~~~~~~~~~~~~~~~~~~~~~~~~~~~~~~~~~~~~~~~~~~~~~~~~~~~~~~\nonumber\\-\frac{2}{3\gamma_c}
\left[1-2\gamma_c+\frac{\alpha
H^2}{3c^2M_P^2}\left(\frac{\Omega_{\Lambda}}{c^2\gamma_c}\right)\right]
\left[1-\left(\frac{\Omega_\Lambda}{c^2\gamma_c}\right)^{1/2}\cos
n\Big(\sqrt{|k|}y\Big)\right].\label{wECHDE}
\end{eqnarray}
Note that as we already mentioned, at the very early stage when the
universe undergoes an inflation phase, the correction terms in the
ECHDE density (\ref{ECHDE}) become important. After the end of the
inflationary phase, the universe subsequently enters in the
radiation and then matter dominated eras. In these two epochs, since
the universe is much larger, the entropy-corrected terms to ECHDE,
namely the last two terms in Eq. (\ref{ECHDE}), can be safely
ignored. Therefore if we set $\alpha=\beta=0$, then from Eq.
(\ref{gamma n}) $\gamma_c=1$ and Eq. (\ref{wECHDE}) recovers the EoS
parameter of the ordinary HDE \cite{BWang}
\begin{eqnarray}
\omega_\Lambda=-\frac{1}{3}-b^2\left(\frac{1+\Omega_k}{\Omega_\Lambda}\right)-\frac{2\sqrt{\Omega_\Lambda}}{3c}\cos
n\Big(\sqrt{|k|}y\Big).
\end{eqnarray}
In next sections, we suggest a correspondence between the
interacting ECHDE model with the quintessence, tachyon, K-essence
and dilaton scalar field models in the non-flat universe.

\section{Entropy-corrected holographic quintessence model}

Quintessence is described by an ordinary time dependent and
homogeneous scalar field $\phi$ which is minimally coupled to
gravity, but with a particular potential $V(\phi)$ that leads to the
accelerating universe. The action for quintessence is given by
\cite{Copeland}
\begin{equation}\label{1}
S=\int {\rm d}^4x\sqrt{-g}\left[
-\frac{1}{2}g^{\mu\nu}\partial_\mu\phi\partial_\nu\phi-V(\phi)
\right].
\end{equation}
The energy momentum tensor of the field is derived by varying the
action (\ref{1}) with respect to $g^{\mu\nu}$:
\begin{equation}\label{2}
T_{\mu\nu}=-\frac{2}{\sqrt{-g}}\frac{\delta S}{\delta g^{\mu\nu}},
\end{equation}
which gives
\begin{equation}\label{3}
T_{\mu\nu}=\partial_\mu\phi\partial_\nu\phi-g_{\mu\nu}
\left[\frac{1}{2}g^{\alpha\beta}\partial_\alpha\phi\partial_\beta\phi
+V(\phi)\right].
\end{equation}
 The energy density and pressure of the quintessence scalar field
$\phi$ are as follows
\begin{equation}
\rho_Q=-T^0_0=\frac{1}{2}\dot \phi^2+V(\phi),\label{ro q}
\end{equation}
\begin{equation}
p_Q=T_i^i=\frac{1}{2}\dot \phi^2-V(\phi).\label{p q}
\end{equation}
The EoS parameter for the quintessence scalar field is given by
\begin{equation}
\omega_Q=\frac{p_Q}{\rho_Q}=\frac{\dot \phi^2-2V(\phi)}{\dot
\phi^2+2V(\phi)}.\label{w q}
\end{equation}
From (\ref{w q}) for $\omega_Q<-1/3$, we find that the universe
accelerates when $\dot \phi^2<V(\phi).$

Here we establish the correspondence between the interacting ECHDE
scenario and the quintessence DE model, then equating Eq. (\ref{w
q}) with the EoS parameter of interacting ECHDE (\ref{wECHDE}),
$\omega_Q=\omega_\Lambda$, and also equating Eq. (\ref{ro q}) with
(\ref{ECHDE}), $\rho_Q=\rho_\Lambda$, we have
\begin{equation} \dot
\phi^2=(1+\omega_\Lambda)\rho_\Lambda,\label{phidot2-2}
\end{equation}
\begin{equation}
V(\phi)=\frac{1}{2}(1-\omega_\Lambda)\rho_\Lambda.\label{Vphi-2}
\end{equation}
Substituting Eqs. (\ref{ECHDE}) and (\ref{wECHDE}) into Eqs.
(\ref{phidot2-2}) and (\ref{Vphi-2}), one can obtain the kinetic
energy term and the quintessence potential energy as follows
\begin{eqnarray}
\dot\phi^2=3M_P^2H^2\Omega_{\Lambda}\left(b^2\Big(\frac{1+\Omega_k}{\Omega_\Lambda}\Big)\right.
~~~~~~~~~~~~~~~~~~~~~~~~~~~~~~~~~~~~~~~~~~~~~~~~~~~~~~~\nonumber\\\left.+\frac{2}{3\gamma_c}
\Big[1-2\gamma_c+\frac{\alpha
H^2}{3c^2M_P^2}\Big(\frac{\Omega_{\Lambda}}{c^2\gamma_c}\Big)\Big]
\Big[1-\Big(\frac{\Omega_\Lambda}{c^2\gamma_c}\Big)^{1/2}\cos
n\Big(\sqrt{|k|}y\Big)\Big]\right),\label{fi dot q}
\end{eqnarray}

\begin{eqnarray}
V(\phi)=3M_P^2H^2\Omega_{\Lambda}\left(1+\frac{b^2}{2}\Big(\frac{1+\Omega_k}{\Omega_\Lambda}\Big)\right.
~~~~~~~~~~~~~~~~~~~~~~~~~~~~~~~~~~~~~~~~~~~~~~~~~~~~~~~\nonumber\\\left.+\frac{1}{3\gamma_c}
\Big[1-2\gamma_c+\frac{\alpha
H^2}{3c^2M_P^2}\Big(\frac{\Omega_{\Lambda}}{c^2\gamma_c}\Big)\Big]
\Big[1-\Big(\frac{\Omega_\Lambda}{c^2\gamma_c}\Big)^{1/2}\cos
n\Big(\sqrt{|k|}y\Big)\Big]\right).\label{pot q}
\end{eqnarray}
From Eqs. (\ref{fi dot q}) one can obtain the evolutionary form of
the quintessence scalar field as
\begin{eqnarray}
\phi(a)-\phi(a_0)=M_P\int_{a_0}^{a}\left(3b^2(1+\Omega_k)\right.
~~~~~~~~~~~~~~~~~~~~~~~~~~~~~~~~~~~~~~~~~~~~~~~~~~~~~~~\nonumber\\\left.+\frac{2\Omega_{\Lambda}}{\gamma_c}\Big[1-2\gamma_c+\frac{\alpha
H^2}{3c^2M_P^2}\Big(\frac{\Omega_{\Lambda}}{c^2\gamma_c}\Big)\Big]
\Big[1-\Big(\frac{\Omega_\Lambda}{c^2\gamma_c}\Big)^{1/2}\cos
n\Big(\sqrt{|k|}y\Big)\Big]\right)^{1/2}\frac{{\rm
d}a}{a},\label{phiQ}
\end{eqnarray}
where $a_0$ is the scale factor at the present time.

The above integral cannot be taken analytically. But during the
early inflation era when the correction terms make sense in the
ECHDE density (\ref{ECHDE}), the Hubble parameter $H$ is constant
and $a=a_0e^{Ht}$. Hence the Hubble horizon $H^{-1}$ and the future
event horizon $R_{\rm h}=a\int_t^\infty \frac{{\rm d} t}{a}$ will
coincide i.e. $R_{\rm h} = H^{-1}=$ const. On the other hand, since
an early inflation era leads to a flat universe we have $L = R_{\rm
h} = H^{-1}=$ const. Also from Eqs. (\ref{HL}) and (\ref{cosn}) we
have $\frac{\Omega_{\Lambda}}{c^2\gamma_c}=1$ and $\cos
n\Big(\sqrt{|k|}y\Big)=1$. Therefore during the early inflation era,
Eq. (\ref{phiQ}) reduces to
\begin{equation}
\phi(a)=\phi(a_0)+\sqrt{3}~bM_P\ln{\left(\frac{a}{a_0}\right)}.
\end{equation}
For the late-time universe, i.e. $\Omega_\Lambda=1$ and
$\Omega_k=0$, the universe becomes large and ECHDE reduces to the
ordinary HDE. In this case $L=R_{\rm h}\neq H^{-1}$ and $H\neq$
const. Now by setting $\gamma_c=1(\alpha=\beta=0)$ and $\cos
n\Big(\sqrt{|k|}y\Big)=1$, the Hubble parameter from Eqs. (\ref{HL})
and (\ref{Ldot}) can be obtained as
\begin{equation}
H=\frac{H_0}{1+\Big(\frac{c-1}{c}\Big)H_0(t-t_0)},\label{H1}
\end{equation}
where $H_0$ is the Hubble parameter at the present time. After
integration of Eq. (\ref{H1}) with respect to $t$, the scale factor
can be obtained as
\begin{equation}
a=a_0\left[1+\left(\frac{c-1}{c}\right)H_0(t-t_0)\right]^{\frac{c}{c-1}}.\label{a}
\end{equation}
Using the above relation, one can rewrite Eq. (\ref{H1}) as
\begin{equation}
H=H_0\Big(\frac{a}{a_0}\Big)^{\frac{1-c}{c}}.\label{H2}
\end{equation}
Finally for the late-time universe, Eq. (\ref{phiQ}) yields
\begin{equation}
\phi(a)=\phi(a_0)+M_P\left[3b^2-2\left(1-\frac{1}{c}\right)\right]^{1/2}\ln{\left(\frac{a}{a_0}\right)}.
\end{equation}
\section{Entropy-corrected holographic tachyon model}
In recent years, a huge interest has been devoted in studying the
inflationary model with the help of tachyon field. The tachyon field
associated with unstable D-branes might be responsible for
cosmological inflation in the early evolution of the universe, due
to tachyon condensation near the top of the effective scalar
potential \cite{23}. Also the tachyonic matter could suggests some
new form of DM at late epoch \cite{24}. The tachyon field has
emerged as a possible source of the DE. A rolling tachyon has an
interesting EoS whose parameter smoothly interpolates between $-1$
and $0$ \cite{gibbons}. This discovery motivated to take DE as the
dynamical quantity, i.e. a variable cosmological constant and model
inflation using tachyons. The effective Lagrangian density of
tachyon matter is given by \cite{Sen}
\begin{equation}
{\mathcal{L}}=-V(\phi)\sqrt{1+\partial_{\mu}\phi
\partial^{\mu}\phi}.
\end{equation}
The energy density and pressure for the tachyon field are as
following \cite{Sen}
\begin{equation}
\rho_{T}=\frac{V(\phi)}{\sqrt{1-\dot{\phi}^{2}}},\label{rhot}
\end{equation}
\begin{equation}
p_{T}=-V(\phi)\sqrt{1-\dot{\phi}^{2}},
\end{equation}
where $V(\phi)$ is the tachyon potential. The EoS parameter for the
tachyon scalar field is obtained as
\begin{equation}
\omega_{T}=\frac{p_{T}}{\rho_{T}}=\dot{\phi}^{2}-1.\label{wt}
\end{equation}
If we establish the correspondence between the ECHDE and tachyon DE,
then equating Eq. (\ref{wt}) with the EoS parameter of interacting
ECHDE (\ref{wECHDE}), $\omega_T=\omega_\Lambda$, and also equating
Eq. (\ref{rhot}) with (\ref{ECHDE}), $\rho_T=\rho_\Lambda$, we
obtain
\begin{equation}
\dot{\phi}^{2}=b^2\Big(\frac{1+\Omega_k}{\Omega_\Lambda}\Big)+\frac{2}{3\gamma_c}
\Big[1-2\gamma_c+\frac{\alpha
H^2}{3c^2M_P^2}\Big(\frac{\Omega_{\Lambda}}{c^2\gamma_c}\Big)\Big]
\Big[1-\Big(\frac{\Omega_\Lambda}{c^2\gamma_c}\Big)^{1/2}\cos
n\Big(\sqrt{|k|}y\Big)\Big],\label{fi dotT}
\end{equation}
\begin{eqnarray}
V(\phi)=3M_P^2H^2\Omega_{\Lambda}\left(1+b^2\Big(\frac{1+\Omega_k}{\Omega_\Lambda}\Big)\right.
~~~~~~~~~~~~~~~~~~~~~~~~~~~~~~~~~~~~~~~~~~~~~~~~~~~~~~~\nonumber\\\left.+\frac{2}{3\gamma_c}
\Big[1-2\gamma_c+\frac{\alpha
H^2}{3c^2M_P^2}\Big(\frac{\Omega_{\Lambda}}{c^2\gamma_c}\Big)\Big]
\Big[1-\Big(\frac{\Omega_\Lambda}{c^2\gamma_c}\Big)^{1/2}\cos
n\Big(\sqrt{|k|}y\Big)\Big]\right)^{1/2}.\label{pot T}
\end{eqnarray}
From Eq. (\ref{fi dotT}), one can obtain the evolutionary form of
the tachyon scalar field as
\begin{eqnarray}
\phi(a)-\phi(a_0)=\int_{a_0}^{a}\frac{{\rm
d}a}{Ha}\left(b^2\Big(\frac{1+\Omega_k}{\Omega_\Lambda}\Big)\right.
~~~~~~~~~~~~~~~~~~~~~~~~~~~~~~~~~~~~~~~~~~~~~~~~~~~~~~~\nonumber\\\left.+\frac{2}{3\gamma_c}
\Big[1-2\gamma_c+\frac{\alpha
H^2}{3c^2M_P^2}\Big(\frac{\Omega_{\Lambda}}{c^2\gamma_c}\Big)\Big]
\Big[1-\Big(\frac{\Omega_\Lambda}{c^2\gamma_c}\Big)^{1/2}\cos
n\Big(\sqrt{|k|}y\Big)\Big]\right)^{1/2}.\label{phiT}
\end{eqnarray}
During the early inflation era ($L = R_{\rm h} = H^{-1}=$ const.),
Eq. (\ref{phiT}) yields
\begin{equation}
\phi(a)=\phi(a_0)+\frac{b}{H\sqrt{\Omega_{\Lambda}}}\ln{\left(\frac{a}{a_0}\right)},
\end{equation}
where
\begin{equation}
\Omega_{\Lambda}=c^2+\frac{H^2}{3M_P^2}\Big[\alpha\ln(M_P^2H^{-2})+\beta\Big].\label{OmegaLambda}
\end{equation}
For the late-time universe, i.e. $\Omega_\Lambda=1$, $\Omega_k=0$
and $\gamma_c=1 (\alpha=\beta=0)$, using Eq. (\ref{H2}) one can take
the integral (\ref{phiT}) as
\begin{equation}
\phi(a)=\phi(a_0)+\frac{\left[b^2-\frac{2}{3}\left(1-\frac{1}{c}\right)\right]^{1/2}}{H_0\left(1-\frac{1}{c}\right)}
\left[\left(\frac{a}{a_0}\right)^{1-\frac{1}{c}}-1\right].
\end{equation}
\section{Entropy-corrected holographic K-essence model}

A model in which the kinetic energy term of the scalar field appears
in the Lagrangian in a non-canonical way is termed the K-essence
model. Such fields were originally used to model inflation, a
scenario called K-inflation \cite{inf}. Moreover the stable tracker
solutions for K-essence have been obtained i.e. solutions which
start from arbitrary initial conditions and reach to the same final
state of cosmic acceleration \cite{att}. The purpose of introducing
K-essence is to provide a dynamical explanation which does not
require the fine-tuning of initial conditions. It is also possible
to have a situation where the accelerated expansion of the universe
arises out of modifications to the kinetic energy of the scalar
fields. The K-essence is described by a general scalar field action
which is a function of $\phi$ and $\chi=\dot{\phi}^2/2$, and is
given by \cite{Chiba, Picon3}
\begin{equation} S=\int {\rm d} ^{4}x\sqrt{-{\rm
g}}~p(\phi,\chi),
\end{equation}
where $p(\phi,\chi)$ corresponds to a pressure density as
\begin{equation}
p(\phi,\chi)=f(\phi)(-\chi+\chi^{2}),
\end{equation}
and the energy density of the field $\phi$ is
\begin{equation}
\rho(\phi,\chi)=f(\phi)(-\chi+3\chi^{2}).\label{rhok}
\end{equation}
The EoS parameter for the K-essence scalar field is obtained as
\begin{equation}
\omega_{K}=\frac{p(\phi,\chi)}{\rho(\phi,\chi)}=\frac{\chi-1}{3\chi-1}.\label{wk}
\end{equation}
Equating Eq. (\ref{wk}) with the EoS parameter (\ref{wECHDE}),
$\omega_{K}=\omega_{\Lambda}$, we find the solution for $\chi$
\begin{equation}
\chi=\frac{2+b^2\Big(\frac{1+\Omega_k}{\Omega_\Lambda}\Big)+\frac{2}{3\gamma_c}
\Big[1-2\gamma_c+\frac{\alpha
H^2}{3c^2M_P^2}\Big(\frac{\Omega_{\Lambda}}{c^2\gamma_c}\Big)\Big]
\Big[1-\Big(\frac{\Omega_\Lambda}{c^2\gamma_c}\Big)^{1/2}\cos
n\Big(\sqrt{|k|}y\Big)\Big]}
{4+3b^2\Big(\frac{1+\Omega_k}{\Omega_\Lambda}\Big)+\frac{2}{\gamma_c}
\Big[1-2\gamma_c+\frac{\alpha
H^2}{3c^2M_P^2}\Big(\frac{\Omega_{\Lambda}}{c^2\gamma_c}\Big)\Big]
\Big[1-\Big(\frac{\Omega_\Lambda}{c^2\gamma_c}\Big)^{1/2}\cos
n\Big(\sqrt{|k|}y\Big)\Big]}.\label{khi}
\end{equation}
Using $\dot{\phi}^2=2\chi$ and (\ref{khi}), we obtain the
evolutionary form of the K-essence scalar field as
\begin{eqnarray}
\phi(a)=\phi(a_0)~~~~~~~~~~~~~~~~~~~~~~~~~~~~~~~~~~~~~~~~~~~~~~~~~~~~~~~~~~~~~~~~~~~~~~~~~~~~~~\nonumber\\+\int_{a_0}^{a}\frac{{\rm
d}a}{Ha}\left(\frac{1+\frac{b^2}{2}(\frac{1+\Omega_k}{\Omega_\Lambda})+\frac{1}{3\gamma_c}
\Big[1-2\gamma_c+\frac{\alpha
H^2}{3c^2M_P^2}\Big(\frac{\Omega_{\Lambda}}{c^2\gamma_c}\Big)\Big]
\Big[1-\Big(\frac{\Omega_\Lambda}{c^2\gamma_c}\Big)^{1/2}\cos
n\Big(\sqrt{|k|}y\Big)\Big]}
{1+\frac{3b^2}{4}(\frac{1+\Omega_k}{\Omega_\Lambda})+\frac{1}{2\gamma_c}
\Big[1-2\gamma_c+\frac{\alpha
H^2}{3c^2M_P^2}\Big(\frac{\Omega_{\Lambda}}{c^2\gamma_c}\Big)\Big]
\Big[1-\Big(\frac{\Omega_\Lambda}{c^2\gamma_c}\Big)^{1/2}\cos
n\Big(\sqrt{|k|}y\Big)\Big]}\right)^{1/2}.\label{phiK}
\end{eqnarray}
During the early inflation era, Eq. (\ref{phiK}) reduces to
\begin{equation}
\phi(a)=\phi(a_0)+\frac{1}{H}\left(\frac{4\Omega_{\Lambda}+2b^2}{4\Omega_{\Lambda}+3b^2}\right)^{1/2}\ln{\left(\frac{a}{a_0}\right)},
\end{equation}
where $\Omega_{\Lambda}$ is given by Eq. (\ref{OmegaLambda}).

For the late-time universe, i.e. $L=R_{\rm h}\neq H^{-1}$ and
$H\neq$ const., using Eq. (\ref{H2}) one can take the integral
(\ref{phiK}) as
\begin{equation}
\phi(a)=\phi(a_0)+\left(\frac{1+\frac{b^2}{2}-\frac{1}{3}(1-\frac{1}{c})}{1+\frac{3b^2}{4}-\frac{1}{2}(1-\frac{1}{c})}\right)^{1/2}
\frac{\left[\left(\frac{a}{a_0}\right)^{1-\frac{1}{c}}-1\right]}{H_0(1-\frac{1}{c})}.
\end{equation}
\section{Entropy-corrected holographic dilaton model}
The process of compactification of the string theory from higher to
four dimensions introduces the scalar dilaton field which is coupled
to curvature invariants. The coefficient of the kinematic term of
the dilaton can be negative in the Einstein frame, which means that
the dilaton behaves as a phantom-type scalar field. The pressure
(Lagrangian) density and the energy density of the dilaton DE model
is given by \cite{Gasperini}
\begin{equation}
p_{D}=-\chi+c'e^{\lambda\phi}\chi^{2},
\end{equation}
\begin{equation}
\rho_{D}=-\chi+3c'e^{\lambda\phi}\chi^{2},\label{rhod}
\end{equation}
where $c'$ and $\lambda$ are positive constants and
$\chi=\dot{\phi}^2/2$. The EoS parameter for the dilaton scalar
field is given by
\begin{equation}
\omega_{D}=\frac{p_D}{\rho_D}=\frac{-1+c'e^{\lambda\phi}\chi}{-1+3c'e^{\lambda\phi}\chi}.\label{wd}
\end{equation}
Equating Eq. (\ref{wd}) with the EoS parameter (\ref{wECHDE}),
$\omega_D=\omega_\Lambda$, we find the following solution
\begin{equation}
c'e^{\lambda\phi}\chi=\frac{2+b^2\Big(\frac{1+\Omega_k}{\Omega_\Lambda}\Big)+\frac{2}{3\gamma_c}
\Big[1-2\gamma_c+\frac{\alpha
H^2}{3c^2M_P^2}\Big(\frac{\Omega_{\Lambda}}{c^2\gamma_c}\Big)\Big]
\Big[1-\Big(\frac{\Omega_\Lambda}{c^2\gamma_c}\Big)^{1/2}\cos
n\Big(\sqrt{|k|}y\Big)\Big]}
{4+3b^2\Big(\frac{1+\Omega_k}{\Omega_\Lambda}\Big)+\frac{2}{\gamma_c}
\Big[1-2\gamma_c+\frac{\alpha
H^2}{3c^2M_P^2}\Big(\frac{\Omega_{\Lambda}}{c^2\gamma_c}\Big)\Big]
\Big[1-\Big(\frac{\Omega_\Lambda}{c^2\gamma_c}\Big)^{1/2}\cos
n\Big(\sqrt{|k|}y\Big)\Big]},\label{khi D}
\end{equation}
then using $\dot{\phi}^2=2\chi$, we obtain
\begin{equation}
e^{\frac{\lambda\phi}{2}}\dot{\phi}=\left(\frac{4+2b^2\Big(\frac{1+\Omega_k}{\Omega_\Lambda}\Big)+\frac{4}{3\gamma_c}
\Big[1-2\gamma_c+\frac{\alpha
H^2}{3c^2M_P^2}\Big(\frac{\Omega_{\Lambda}}{c^2\gamma_c}\Big)\Big]
\Big[1-\Big(\frac{\Omega_\Lambda}{c^2\gamma_c}\Big)^{1/2}\cos
n\Big(\sqrt{|k|}y\Big)\Big]}
{c'\Big(4+3b^2\Big(\frac{1+\Omega_k}{\Omega_\Lambda}\Big)+\frac{2}{\gamma_c}
\Big[1-2\gamma_c+\frac{\alpha
H^2}{3c^2M_P^2}\Big(\frac{\Omega_{\Lambda}}{c^2\gamma_c}\Big)\Big]
\Big[1-\Big(\frac{\Omega_\Lambda}{c^2\gamma_c}\Big)^{1/2}\cos
n\Big(\sqrt{|k|}y\Big)\Big]\Big)}\right)^{1/2}.
\end{equation}
Integrating with respect to $a$, we get
\begin{eqnarray}
e^\frac{\lambda\phi(a)}{2}=e^\frac{\lambda\phi(a_0)}{2}
~~~~~~~~~~~~~~~~~~~~~~~~~~~~~~~~~~~~~~~~~~~~~~~~~~~~~~~~~~~~~~~~~~~~~~~~~~~~~~\nonumber\\+\frac{\lambda}{2\sqrt{c'}}\int_{a_0}^{a}\frac{{\rm
d}a}{Ha}\left(\frac{4+2b^2\Big(\frac{1+\Omega_k}{\Omega_\Lambda}\Big)+\frac{4}{3\gamma_c}
\Big[1-2\gamma_c+\frac{\alpha
H^2}{3c^2M_P^2}\Big(\frac{\Omega_{\Lambda}}{c^2\gamma_c}\Big)\Big]
\Big[1-\Big(\frac{\Omega_\Lambda}{c^2\gamma_c}\Big)^{1/2}\cos
n\Big(\sqrt{|k|}y\Big)\Big]}
{4+3b^2\Big(\frac{1+\Omega_k}{\Omega_\Lambda}\Big)+\frac{2}{\gamma_c}
\Big[1-2\gamma_c+\frac{\alpha
H^2}{3c^2M_P^2}\Big(\frac{\Omega_{\Lambda}}{c^2\gamma_c}\Big)\Big]
\Big[1-\Big(\frac{\Omega_\Lambda}{c^2\gamma_c}\Big)^{1/2}\cos
n\Big(\sqrt{|k|}y\Big)\Big]}\right)^{1/2}.
\end{eqnarray}
Therefore the evolutionary form of the dilaton scalar field is
obtained as
\begin{eqnarray}
\phi(a)=\frac{2}{\lambda}\ln\left[e^\frac{\lambda\phi(a_0)}{2}\right.
~~~~~~~~~~~~~~~~~~~~~~~~~~~~~~~~~~~~~~~~~~~~~~~~~~~~~~~~~~~~~~~~~~~~~~~~~~~~~~\nonumber\\\left.+\frac{\lambda}{\sqrt{2c'}}\int_{a_0}^{a}\frac{{\rm
d}a}{Ha}\left(\frac{2+b^2\Big(\frac{1+\Omega_k}{\Omega_\Lambda}\Big)+\frac{2}{3\gamma_c}
\Big[1-2\gamma_c+\frac{\alpha
H^2}{3c^2M_P^2}\Big(\frac{\Omega_{\Lambda}}{c^2\gamma_c}\Big)\Big]
\Big[1-\Big(\frac{\Omega_\Lambda}{c^2\gamma_c}\Big)^{1/2}\cos
n\Big(\sqrt{|k|}y\Big)\Big]}
{4+3b^2\Big(\frac{1+\Omega_k}{\Omega_\Lambda}\Big)+\frac{2}{\gamma_c}
\Big[1-2\gamma_c+\frac{\alpha
H^2}{3c^2M_P^2}\Big(\frac{\Omega_{\Lambda}}{c^2\gamma_c}\Big)\Big]
\Big[1-\Big(\frac{\Omega_\Lambda}{c^2\gamma_c}\Big)^{1/2}\cos
n\Big(\sqrt{|k|}y\Big)\Big]}\right)^{1/2}\right].\label{phiD}
\end{eqnarray}
During the early inflation era, Eq. (\ref{phiD}) yields
\begin{equation}
\phi(a)=\frac{2}{\lambda}\ln\left[e^{\frac{\lambda\phi(a_0)}{2}}+\frac{\lambda}{\sqrt{2c'}}\frac{1}{H}
\left(\frac{2\Omega_{\Lambda}+b^2}{4\Omega_{\Lambda}+3b^2}\right)^{1/2}\ln{\left(\frac{a}{a_0}\right)}\right],
\end{equation}
where $\Omega_{\Lambda}$ is given by Eq. (\ref{OmegaLambda}).

For the late-time universe, using Eq. (\ref{H2}) one can take the
integral (\ref{phiD}) as
\begin{equation}
\phi(a)=\frac{2}{\lambda}\ln\left\{e^{\frac{\lambda\phi(a_0)}{2}}+\frac{\lambda}{\sqrt{2c'}}
\left(\frac{2+b^2-\frac{2}{3}\Big(1-\frac{1}{c}\Big)}{4+3b^2-2\Big(1-\frac{1}{c}\Big)}\right)^{1/2}
\frac{\left[\left(\frac{a}{a_0}\right)^{1-\frac{1}{c}}-1\right]}{H_0(1-\frac{1}{c})}\right\}.
\end{equation}
\section{Conclusions}

Here we considered the entropy-corrected version of the HDE model
which is in interaction with DM in the non-flat FRW universe. The
HDE model is an attempt for probing the nature of DE within the
framework of quantum gravity \cite{Guberina1}. We considered the
logarithmic correction term to the energy density of HDE model.
The addition of correction terms to the energy density of HDE is
motivated from the LQG which is one of the promising theories of
quantum gravity. Using this modified energy density, we obtained
the EoS parameter for the interacting ECHDE. We established a
correspondence between the interacting ECHDE model with the
quintessence, tachyon, K-essence and dilaton scalar field models
in the non-flat FRW universe. These correspondences are important
to understand how various candidates of DE are mutually related to
each other.

In the present case, the correspondence is established between ECHDE
and various scalar field models of DE. We adopted the viewpoint that
these scalar field models of DE are effective theories of an
underlying theory of DE. Thus, we should be capable of using these
scalar field models to mimic the evolving behavior of the
interacting ECHDE and reconstructing the scalar field models
according to the evolutionary behavior of the interacting ECHDE. We
reconstructed the potentials and the dynamics of these scalar field
models, which describe accelerated expansion of the universe,
according to the evolutionary behavior of the interacting ECHDE
model. We also obtained the explicit evolutionary forms of the
corresponding scalar fields for the both of early inflation
($L=R_{\rm h}=H^{-1}$= const.) and late-time acceleration ($L=R_{\rm
h}\neq H^{-1}$ and $H\neq$ const.) phases. For late-time
acceleration, $L$ is dynamical and $\dot L$ will contribute in the
above expressions and will yield the scalar potentials for the DE.
This suggests that the vacuum energy that produced inflation at
early cosmic epoch and the one driving late-time cosmic acceleration
are fundamentally different. Hence the same scalar field moves in
different potentials at different times.
\\
\\
\noindent{{\bf Acknowledgements}}\\
The authors thank the reviewers for valuable comments. The work of
K. Karami has been supported financially by Research Institute for
Astronomy $\&$ Astrophysics of Maragha (RIAAM), Maragha, Iran.

\end{document}